\documentclass[11pt,fleqn]{article}

\usepackage{amsmath,amssymb}
\usepackage{graphicx}
\usepackage[top=0.75in, bottom=0.75in, left=0.75in, right=0.75in, dvips]{geometry}

\begin{document}

\title{Massless Scalar Field Propagator in a Quantized Space-Time}

\author{V.\ Elias\thanks{Department of Applied Mathematics, The University of Western Ontario, London, Ontario  N6A 5B7, Canada} ~and
T.G.\ Steele\thanks{Department of Physics and Engineering Physics, University of
Saskatchewan, Saskatoon, SK, S7N 5E2, Canada}
}

\maketitle

\begin{abstract}
We consider in detail the analytic behaviour of the non-interacting massless scalar field two-point function in H.S.~Snyder's discretized non-commuting spacetime.  The propagator we find is purely real on the Euclidean side of the complex $p^2$ plane and goes like  $1/p^2$  as $p^2\to 0$ from either the Euclidean or Minkowski side.  The real part of the propagator goes smoothly to zero as $p^2$ increases to the discretization scale $1/a^2$ {\em and remains zero} for $p^2>1/a^2$.  This behaviour is consistent with the termination of single-particle propagation  on the ultraviolet side of the discretization scale.  The imaginary part of the propagator, consistent with a multiparticle-production branch discontinuity, is finite and continuous on the Minkowski side, slowly falling to zero when $1/a^2<p^2<\infty$.  Finally, we argue that the spectral function for the multiparticle states appears to saturate as $p^2$ probes just beyond the $1/a^2$ discretization scale.  We speculate on the cosmological consequences of such a spectral function.   
\end{abstract}

Some years ago, J.C.\ Breckenridge, T.G.\ Steele and V.\ Elias \cite{1} examined  quantum field theory
for massless scalar $\phi^4$ theory in a flat quantised space-time that was first put forward by Snyder \cite{2}.
The Snyder algebra \cite{2} introduces a spacetime discretization scale $a$ through a non-zero commutator of spacetime operators:
\begin{gather}
\left[ x^\mu, \; x^\nu\right]  =  ia^2 J^{\mu\nu}
\label{st_comm}
\\
\left[ x^\mu, \; J^{\rho\sigma} \right]  =  i \left( g^{\mu \rho}x^\sigma - g^{\mu\sigma} x^\rho\right)
\\
\left[ J_{\mu\nu}, \; J_{\rho\sigma} \right]  =  i \left[ g_{\nu\rho} J_{\mu\sigma} - g_{\mu\rho} J_{\nu\sigma} + g_{\mu\sigma} J_{\nu\rho} - g_{\nu\sigma} J_{\mu\rho} \right]
\label{eq1}
\end{gather}
where $J_{\mu\nu}=-J_{\nu\mu}$ corresponds to generators of Lorentz transformations  (rotations $L_i = \epsilon_{ijk} (J_{jk})/2$, and  boosts $M_j = J_{j0}$). 
A consequence of the algebra (\ref{st_comm}) is the quantization of a spatial coordinate in multiples of $a$ \cite{2}.  We emphasize that this is more complex than a lattice formulation of spacetime structure; the non-zero commutators of spatial operators implies that different coordinates are not simultaneously observable, analogous to quantization of angular momentum in quantum mechanics.  Furthermore, the temporal operator results in a continuous spectrum \cite{2}, a result that is best illuminated 
by the formulation of the Snyder algebra of non-commuting space-time coordinates in a 5-dimensional deSitter space that is isomorphic to the Poincar\'e group in the limit of vanishing curvature.

In Snyder's formulation, the generators of the Poincar\'e group satisfy their usual algebra
\begin{gather}
\left[p_\mu~,~p_\nu\right]=0
\\
\left[p_\mu~,~J_{\rho\sigma}\right]=i\left( g_{\mu\rho}p_\sigma -g_{\mu\sigma}p_\rho\right)
\end{gather}
and hence four-momentum retains its continuous spectrum.  However, the spacetime commutation relation 
(\ref{st_comm}) introduces the discretization scale $a$ into the commutator $\left[x^\mu~,~p^\nu\right]$
\begin{equation}
\left[x_\mu~,~p^\nu\right]=i\left(\delta_\mu^\nu-a^2p_\mu p^\nu\right)~.
\label{xp_comm}
\end{equation}
A consequence of (\ref{xp_comm}) and the continuous spectrum of the momentum operators is the operator 
relation \cite{2}
\begin{equation}
x_\mu=i\left(\frac{\partial}{\partial p^\mu} -a^2p_\mu p^\rho\frac{\partial}{\partial p^\rho} \right)~,
\end{equation}
which implies that it is convenient to express the generator of Lorentz transformations as
\begin{equation}
J_{\mu\nu}=p_\mu x_\nu-p_\nu x_\mu ~.
\end{equation}
It is evident from these results that the usual reciprocity between configuration and momentum space is broken by the Snyder algebra.  

Along with the string theory where there exists a minimum observable length scale, the Snyder algebra is frequently cited as one of the motivations for the study of quantum field theory in non-commutative spacetimes.   However, the vast array of techniques that have been developed for such purposes typically assume  $\left[ x^\mu~,x^\nu\right]\sim \theta^{\mu\nu}$, where $\theta^{\mu\nu}$ is a ``constant" operator that has trivial commutation relations with all other operators in the theory \cite{non_comm}.  Clearly this  is not the case for the Snyder algebra, and other approaches to formulating quantum field theory in Snyder space are required.

In Ref.\ \cite{1} massless scalar field theory was constructed in Snyder space while still maintaining continuous momentum space variables, an approach very different from that of prior researchers attempting to restore the symmetry between coordinate and momentum space variables \cite{3,4}.  The underlying principle in this construction is based on the observation that the light-cone operator $x^2$ in the Snyder algebra retains its Lorentz invariant nature
\begin{equation}
\left[ x^2~,~J_{\mu\nu}\right]=0~,
\end{equation}
and hence the two-point Green functions can be constrained to have support on the light cone, just  as occurs in a continuous spacetime ({\it i.e.,} the $a\to 0$ limit.) \cite{1}.

As noted above, fundamental (small) length scales characterise both string theories and quantum gravity.  Hence the construction of a quantum field theory in a quantized space-time in which such a fundamental scale explicitly appears is itself important.  Indeed, Snyder's primary motivation was to find a means to regulate quantum-field-theoretical UV divergences through the self-consistent introduction of a fundamental length scale \cite{2}. In Ref.\ \cite{1}, massless scalar $\lambda \phi^4$ theory was formulated in Snyder's quantized spacetime, and it was demonstrated that the one-loop Green two- and four-point functions were free of both UV and IR divergences. However, 
the massless scalar field propagator calculated in \cite{1} was ambiguous except for Euclidean momenta.  This ambiguity was not an issue for loop calculations (which fundamentally occur in a Euclideanized momentum space).   We have chosen here to revisit this calculation in detail so as to establish the massless scalar field propagator over all real momenta, Euclidean and Minkowskian, because of the relation between particle creation and the analytic properties of the propagator for Minkowskian momentum.

Let us begin with the differential equation (3.13) in Ref.\ \cite{1} for the massless scalar field momentum-space propagator $G(y)$
\begin{equation}
\left[ 2y(1-y) \frac{d}{dy} - (3y-4) \right] G^\prime (y) = 0~,
\label{eq2}
\end{equation}
with $y = a^2 p^2$, and with $a$ being the fundamental length characterising the Snyder algebra (\ref{st_comm}).  The quantity $G(y)$ is related to the two-point function via 
\begin{equation}
\left\langle T\left[ \phi(x) \phi(0)\right]\right\rangle=\int\left[d\mu_p\right]
\langle x\vert p\rangle G(y)~,
\end{equation}
where $\left[d\mu_p\right]$ is the appropriate integration measure in the (continuous) momentum 
space \cite{3}.
It is straightforward to find the solution
\begin{equation}
G^\prime (y) = C \sqrt{|1 - y |} / y^2
\label{eq3}
\end{equation}
to Eq.\ (\ref{eq2}), where $C$ is an as-yet arbitrary constant of integration.  This solution has discontinuities at $y = 1$ and $y = 0$, and is continuous on the $y < 0$ Euclidean side as well as on the $y > 1$ Minkowskian side.  For the Euclidean side, let $z = -a^2 p^2$, where $p^2 < 0$ [{\it i.e.,} where $z = -y$ is positive].  We then find from Eq.\ (\ref{eq3}) that
\begin{equation}
G(-z)  =  -C \int \frac{\sqrt{1+z}}{z^2} dz
 =  \frac{C}{2} \log \left( \frac{\sqrt{1+z} + 1}{\sqrt{1+z} - 1} \right) + \frac{C\sqrt{1+z}}{z} + K~,
\label{eq4}
\end{equation} 
$(z > 0)$, where $C$ is the constant of integration in Eq.\  (\ref{eq3}), and where $K$ is a new constant of integration.

We first note that the physical propagator must vanish at $z \rightarrow \infty$, in which case $K = 0$.  We also note from \cite{1} that the propagator $G(y)$ [or $G(-z)$] must exhibit $1 / p^2$ behaviour as $p^2 \rightarrow 0$, {\it i.e.,} as the four-momentum $|p^2|$ becomes small in magnitude relative to any discretization scale $1 / a^2$:
\begin{equation}
\lim_{z \rightarrow 0^+} z \; G(-z) = C = -a^2 \; .
\label{eq5}
\end{equation}
Hence the scalar field propagator for Euclidean momenta $p^2 < 0$ is completely determined to be
\begin{equation}
G(-z)  =  -\frac{a^2}{2} \left[ \log \left[ \frac{\sqrt{1+z} + 1}{\sqrt{1+z} -1} \right] + 2 \; \frac{\sqrt{1+z}}{z} \right]\quad ;\quad
z  =  -a^2 p^2, \; \; p^2 < 0 ~.
\label{eq6}
\end{equation}
This solution is indeed the one obtained in Ref.\ \cite{1}, and it is this Euclidean form that is required for loop calculations. 

To find a solution for physical momenta $0 < p^2 < 1/a^2$, one must continue Eq.\  (\ref{eq6}) to the region $0 < p^2 < 1/a^2$. As momenta in this region increase, the propagator $G$ begins to probe the discretization scale $a$. Continuing Eq.\  (\ref{eq6}) to
the region $0 < y < 1$, where $y = a^2p^2$, we find that
\begin{equation}
G(y) = -\frac{a^2}{2} \left[ \log \left[ \frac{(\sqrt{1-y} + 1)}{(\sqrt{1-y}-1)} \right] - \frac{2\sqrt{1-y}}{y} \right]~.
\label{eq7}
\end{equation}
As before, we verify that $G(y) \rightarrow 1/p^2$ as $p^2 \rightarrow 0^+$:
\begin{equation}
\lim_{y \rightarrow 0^+} y \; G(y) = \lim_{y \rightarrow 0^+} \left\{ -\frac{a^2}{2} \left[ 2y \log 2 - y \log y - i\pi y - 2\right]\right\} = a^2 \; .
\label{eq8}
\end{equation}
Since $y = a^2 p^2$, $G(y) \rightarrow 1/p^2$ as $p^2 \rightarrow 0^+$.  For the region $0 < p^2 < 1/a^2$, we let $y = 1-r$, where $r$ is real, positive, and less than 1.  Then above the branch cut from $p^2 = 0$ to $p^2 = 1$ we find that $\sqrt{1-y} - 1 = \sqrt{r} - 1 = e^{i\pi} (1 - \sqrt{r})$, and that
\begin{equation}
G(y) = \frac{i\pi a^2}{2} - \frac{a^2}{2} \log \left[ \frac{1 + \sqrt{1-y}}{1 - \sqrt{1-y}}\right] + a^2 \frac{\sqrt{1-y}}{y} \; ; y=-a^2p^2,~0<y\le 1 ~.
\label{eq9}
\end{equation}
The $a^2 \sqrt{1-y} / y$ piece is, of course, just the source of the $1/p^2$ behaviour at the origin.  What is interesting is that the real part of the propagator falls smoothly to zero as $p^2 \rightarrow \left( \frac{1}{a^2}\right)^-$, {\it i.e.,} as $y \rightarrow 1$ from below, and that the imaginary part of the propagator $\left( \pi a^2 / 2\right)$ is constant over the entire $0 < p^2 < 1/a^2$ region. The above statements hold provided $a$ has {\it any} nonzero value.  As the discretization scale $a$ goes to zero, the propagator 
$G \rightarrow a^2 / y = 1/p^2$, which is just the massless scalar field propagator on a {\it continuous} spacetime. Because of its imaginary part when $a \neq 0$, there is a branch discontinuity indicative of a multiparticle state.

Finally, we note if $p^2 > 1/a^2$ $(y > 1)$, that $G(y)$ becomes a purely imaginary quantity. To see this, we first find the most general solution to Eq.\  (\ref{eq3}), with $|1 - y| = y - 1$ and with a new constants of integration $D$ and $E$: 
\begin{equation}
G(y) = D \int \left( \sqrt{y-1} \; / y^2 \right) dy = D \left[ \tan^{-1} \sqrt{y-1} - \sqrt{y-1} / y \right] + E 
\label{eq10}
\end{equation}
For the propagator to vanish at infinity, we see that
\begin{equation}
\lim_{y \rightarrow \infty} G(y) = \frac{\pi}{2} D + E = 0 \; .
\label{eq11}
\end{equation}
We also see that as $y \rightarrow 1^+$
\begin{equation}
\lim_{y \rightarrow 1^+} G(y) = E ~.
\label{eq12}
\end{equation}
Since Eq.\ (\ref{eq9}) shows that
\begin{equation}
\lim_{y \rightarrow 1^-} G(y) = \frac{i \pi a^2}{2} \; ,
\label{eq13}
\end{equation}
continuity of the propagator $G(y)$ at $y = 1$ implies that $E = i \pi a^2 / 2$, in which case $D = -i a^2$ [Eq.\  (\ref{eq11})].
Thus,
\begin{equation}
G(y) = -i a^2 \left[ \tan^{-1} \sqrt{y-1} - \frac{\sqrt{y-1}}{y} \right] + i\pi a^2 / 2, \; \; \; y > 1 \; .
\label{eq14}
\end{equation}
Equations (\ref{eq6}), (\ref{eq9}) and (\ref{eq14}) give the full massless scalar-field propagator over the entire real range of $y = a^2 p^2$.  

\begin{figure}[hbt]
\centering
\includegraphics[scale=0.7]{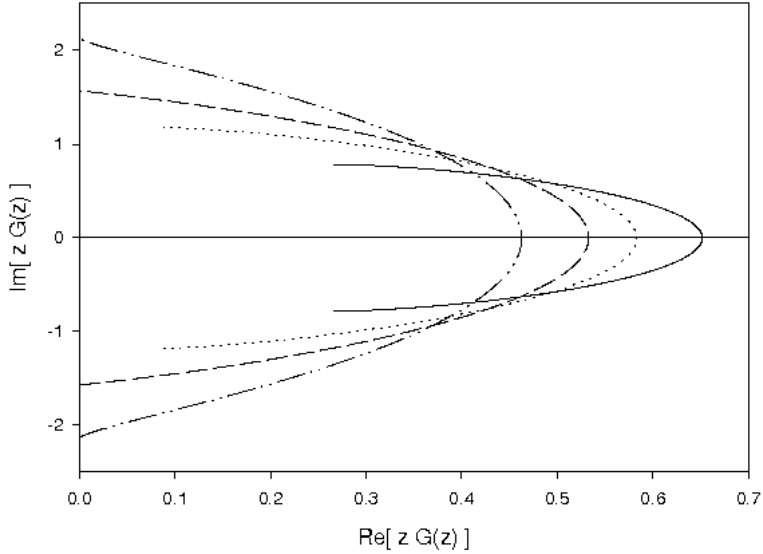}
\caption{Phase plot of $zG(z)$ evaluated around circles of radius $r=\left| a^2p^2\right|$ which begins below the Minkowski branch (lower left end of the curves) and terminates above the Minkowski branch (upper left end of the curves). The solid curve represents $r=1$, dotted curve represents $r=0.75$, dashed curve represents $r=1$, and the dashed-dotted curve  represents $r=1.5$.
}
\label{fig1}
\end{figure}

Salient features of this propagator are:

\begin{itemize}
\item [a)]The Euclidean space propagator is real and continuous.

\item[b)]The real part of the Minkowski space propagator is a continuous function which vanishes at the discretization momentum scale $p^2 = 1/a^2$.

\item[c)]If the discretization scale is exactly zero, the massless scalar field propagator is just $1/p^2$, as we would expect for correspondence to a non-discretized flat space-time.

\item[d)]If the discretization scale is nonzero, the Minkowski space propagator has a continuous imaginary part equal to a constant contribution $\pi a^2 / 2$ for $0 < p^2 \leq 1/a^2$, and then tapering off to zero as $p^2$ increases from $1/a^2$ to $\infty$.

\item[e)]The propagator {\em has no real part at all} for $p^2$ greater than the discretization-scale momentum $1/a^2$.
\end{itemize}
Indeed, the continuity of $Re[G^+(y)]$ and $Im[G^+(y)]$ at $y=1$ is consistent with the result that no discontinuities can occur except across the branch cut.
We note that the results (\ref{eq9}) and (\ref{eq14}) are in complete agreement with the analytic continuation of the Euclidean solution (\ref{eq6}) through the complex plane.  Fig.~\ref{fig1}
shows the phase plot of $zG(z)$ as $z$ traverses a circle of radius $r=\vert z\vert=\left\vert a^2p^2\right\vert$ in the complex plane that terminates on either side of the Minkowskian momentum branch cut.  From the Figure one can observe that the (imaginary part) discontinuity occurs at $Re\left[G(z)\right]=0$ for $p^2>1/a^2$ as discussed above.  As a function of Minkowskian momentum $p^2$, the upper branch of this analytic continuation is in complete numerical agreement with expressions
(\ref{eq9}) and (\ref{eq14}).

Because of its imaginary part over the entire Minkowskian real axis, the propagator necessarily exhibits a branch discontinuity over the entire real $p^2$ axis, indicative of a multiparticle state.\footnote{Weinberg [Ref.\ \protect\cite{5}, p.\ 461] makes the point that 
composite particles may be identified as those whose 
fields do not appear in the Lagrangian.}  
This multiparticle contribution to the K\"allen-Lehman spectral function is just the branch discontinuity itself \cite{5,6}

\begin{equation}
\begin{split}
\sigma(p^2)  = & \frac{1}{2\pi i}\left[ G^+\left(a^2p^2\right)- G^-\left(a^2p^2\right)\right]
=\frac{1}{\pi} Im~ G(a^2 p^2)\nonumber\\
 = & 
\begin{cases}
\frac{a^2}{2} - \frac{a^2}{\pi} \tan^{-1} \sqrt{a^2 p^2 - 1} + \frac{ \sqrt{a^2 p^2-1}}{p^2}~,~ & p^2 > 1/a^2\\ 
\frac{a^2}{2}~,~ & p^2 < 1/a^2
\end{cases}
\end{split}
\end{equation}
where $G^+$ ($G^-$) is the two-point  function above (below) the branch cut on the positive real axis.
The spectral function $\sigma(p^2)$, like $Im G$ itself, is positive definite and falls to zero as $p^2 \rightarrow \infty$.  However, this spectral function, when integrated over $p^2$, is supposed to be bounded from above by 1 \cite{5}.  We find that

\begin{equation}
\int_0^{R/a^2} \sigma(p^2) dp^2  =  
\begin{cases}
\frac{R}{2}~,~ \; \; &R < 1\\
   \frac{R}{2} - \frac{1}{\pi} (R-1) \tan^{-1} (\sqrt{R-1})
 +  \frac{3}{\pi} \sqrt{R-1} - \frac{3}{\pi} \tan^{-1} (\sqrt{R-1})~, \; \; &R > 1
\end{cases}
\label{eq25}
\end{equation}

One sees from (\ref{eq25}) that saturation of the multiparticle spectral function occurs shortly past $p^2 = 1/a^2$;  the integral (\ref{eq25}) equals unity when $R \cong 1.2$.  One also sees runaway behaviour of the spectral function integral as $R \rightarrow \infty$

\begin{equation}
\int_0^{R/a^2} \sigma (p^2) dp^2 
\begin{array}{c}{} \\ \longrightarrow \\ ^{R \rightarrow \infty}
\end{array}
\frac{3}{\pi} \sqrt{R-1} - 1 + {\cal{O}} \left( \frac{1}{\sqrt{R}}\right),
\end{equation}
suggesting that the massless scalar field in a discretized space time is characterized by runaway multiparticle states for $p^2$ greater than the $1/a^2$ discretization scale.  

One can only speculate on cosmological implications of the transfer of energy from scalar fields in a quantized spacetime
via multi-particle creation. All available energy from the massless scalar field would be transferred once $p^2 \sim 1/a^2$, corresponding to saturation of the spectral-function integral.  Indeed, the vanishing of the real part of the propagator at $p^2 = 1/a^2$ may be seen to represent the disappearance of the original single-particle state once the $1/a^2$ discretization scale has been reached.

We are grateful for discussions with A.~Buchel, R.~Dick, G.~McKeon and V.~Miransky, and for financial support from the Natural Sciences and Engineering Research Council of Canada (NSERC).

\end{document}